# Strong dispersive coupling of a high finesse cavity to a micromechanical membrane

J. D. Thompson[1], B. M. Zwickl[1], A. M. Jayich[1], Florian Marquardt[2], S. M. Girvin[1,3], & J. G. E. Harris[1,3]

[1] *Department of Physics, Yale University, 217 Prospect Street, New Haven CT, 06520, USA*

[2] *Physics Department, Center for NanoScience, and Arnold Sommerfeld Center for Theoretical Physics, Ludwig Maximillians University, Theresienstrasse 37, 80333, Munich, Germany*

[3] *Department of Applied Physics, Yale University, 15 Prospect Street, New Haven CT, 06520, USA*

**Macroscopic mechanical objects and electromagnetic degrees of freedom can couple to each other via radiation pressure. Optomechanical systems with sufficiently strong coupling are predicted to exhibit quantum effects and are a topic of considerable interest. Devices which reach this regime would offer new types of control over the quantum state of both light and matter[1,2,3,4,5], and would provide a new arena in which to explore the boundary between quantum and classical physics[6,7,8]. Experiments to date have achieved sufficient optomechanical coupling to laser-cool mechanical devices[9,10,11,12,13,14], but have not yet reached the quantum regime. The outstanding technical challenge in this field is integrating sensitive micromechanical elements (which must be small, light, and flexible) into high finesse cavities (which are typically much more rigid and massive) without compromising the mechanical or optical properties of either. A second, and more fundamental, challenge is to read out the mechanical element's quantum state: displacement**

**measurements (no matter how sensitive) cannot determine the energy eigenstate of an oscillator[15], and measurements which couple to quantities other than displacement[16,17,18] have been difficult to realize in practice. Here we present a novel optomechanical system which seems to resolve both these challenges. We demonstrate a cavity which is detuned by the motion of a 50 nm thick dielectric membrane placed between two macroscopic, rigid, high-finesse mirrors. This approach segregates optical and mechanical functionality to physically distinct structures and avoids compromising either. It also allows for direct measurement of the square of the membrane's displacement, and thus in principle the membrane's energy eigenstate. We estimate it should be practical to use this scheme to observe quantum jumps of a mechanical system, a major goal in the field of quantum measurement.**

Experiments and theoretical proposals aiming to study quantum aspects of the interaction between optical cavities and mechanical objects have focused on cavities in which one of the mirrors defining the cavity is free to move (e.g., in response to radiation pressure exerted by light in the cavity). A schematic of such a setup is shown in Fig.1 (a). Although quite simple, this schematic captures the relevant features of nearly all optomechanical systems described in the literature, including cavities with "folded" geometries, cavities in which multiple mirrors are free to move[5], and whispering gallery mode resonators[14] (WGMRs) in which light is confined to a waveguide by total internal reflection. All these approaches share two important features with the device illustrated in Fig. 1(a). First, the optical cavity's detuning is proportional to the displacement of a mechanical degree of freedom (i.e., mirror displacement or waveguide elongation). Second, a single device is responsible for providing both optical confinement and mechanical pliability.

These systems have realized sufficiently strong optomechanical coupling to laser cool their Brownian motion by a factor of 400 via passive cooling[13]. However the coupling has been insufficient to observe quantum effects such as quantum fluctuations (shot noise) of the radiation pressure. To illustrate the connection between the challenge of observing quantum effects and the properties of the devices illustrated in Fig. 1(a) we consider a figure of merit $R$, the ratio between the force power spectral densities (PSDs) of radiation pressure shot noise $S_F^{(\gamma)}$ and thermal fluctuations $S_F^{(T)}$: $R \equiv S_F^{(\gamma)}/S_F^{(T)} = 16\hbar P_{in} Q F^2 / \lambda c \pi k_B T m \omega_m$. Here $P_{in}$ and $\lambda$ are the laser power and wavelength incident on the cavity; $F$ is the cavity finesse; $Q$, $m$, and $\omega_m$ are the mechanical element's quality factor, motional mass, and resonant frequency; and $T$ is the temperature of the thermal bath. This expression highlights the importance of achieving good optical properties (high $F$) and good mechanical properties (high $Q$; small $m$, $k$) simultaneously.

Simultaneously achieving good mechanical and optical properties has been the main technical barrier to realizing quantum optomechanical systems. In large part this is because high-finesse mirrors are not easily integrated into micromachined devices. These mirrors are typically $SiO_2/Ta_2O_5$ multilayers which are mechanically lossy[19] (limiting $Q$); they must also be ~2 μm thick and ~30 μm in diameter to avoid transmission and diffraction losses[20,21] (setting lower limits on $m$ and $k$); and the mirror's cleanliness and flatness must be maintained during micromachining. As a result most experiments (including those using WGMRs) realize a compromise between high quality optical or mechanical properties.

Figure 1(b) shows a cavity layout which is different from Fig. 1(a) and is the focus of this paper. The cavity is a standard high-finesse Fabry-Perot which in our lab is formed between two macroscopic, rigid, commercial mirrors mounted to a rigid Invar spacer. These mirrors are assumed to be fixed. The mechanically compliant element is a thin dielectric membrane (shown in blue) placed at the waist of the cavity mode. We use a commercial SiN membrane 1 mm square and 50 nm thick. The membrane is supported by a silicon frame (a typical device is shown in Fig. 1(c)). The cavity is excited by a cw Nd:YAG laser ($\lambda$ = 1064 nm); the beam path is shown in Fig. 1(d).

Unlike the cavity type illustrated in Fig. 1(a), the coupling between the membrane and the optical cavity depends upon where the membrane is placed relative to the nodes and antinodes of the cavity mode (shown in green in Fig. 1(b)). This results in a cavity detuning $\omega_{cav}(x)$ which is a periodic function of the membrane displacement $x$, in analogy with the dispersive coupling in some atom-cavity experiments[22,23]. A 1-D calculation gives $\omega_{cav}(x) = (c/L)\cos^{-1}(r_c\cos(4\pi x/\lambda))$ where $L$ is the cavity length and $r_c$ is the (field) reflectivity of the membrane. Figure 1(e) shows a plot of $\omega_{cav}(x)$ for various values of $r_c$. We note that Ref. 24 discusses this geometry, though not its connection to the fabrication and quantum nondemolition issues discussed here.

The optical force on the membrane is $\propto \partial \omega_{cav}/\partial x \propto r_c$, so using a membrane with modest $r_c$ does not substantially reduce the optomechanical coupling. Thus our approach removes the need to integrate good mirrors into good mechanical devices, and thereby avoids the compromises described above. We have exploited the fact that the cavity mirrors set $F$, while the mechanical element's reflectivity only determines the fraction of intracavity photons which transfer momentum to the membrane.

For this "membrane-in-the-middle" approach to work, the membrane must not diminish $F$ through absorption, scatter, or coupling light into lower-$F$ cavity modes. We measured the cavity finesse with the membrane removed ($F_0$) to determine the mirrors' quality, and then with the membrane in place ($F_M$) to determine the loss it introduces. Fig. 2(a) shows the cavity ringdown without (red) and with (blue) the membrane. The laser is switched off at $t$ = 400 ns using an AOM. The fits yield $F_0$ = 16,100 and $F_M$ = 15,200. This difference would imply a total membrane-related optical loss $\beta_M$ = 1.2 x $10^{-5}$. However this is consistent with the variation in $F_0$ produced by slight changes of the cavity mode. In fact some measurements show $F_M > F_0$ so we interpret this value of $\beta_M$ as an upper limit on membrane-induced optical loss. We have found $\beta_M$ is sensitive to the membrane's angular alignment, suggesting it could be minimized with *in situ* adjustments.

Figure 2(b) shows the transmission through the cavity as function of laser frequency and the position of the membrane. The bright bands in the data correspond to the cavity resonances; fitting the data gives $r_c = 0.31$ for this membrane.

Figure 2(c) shows the ringdown of the membrane's lowest flexural resonance at $\omega_m = 2\pi \times 134$ kHz. $m$ is calculated to be $4 \times 10^{-8}$ g, giving a spring constant $k = 28$ N/m. Fitting the data in Fig. 2(c) gives $Q = 1.1 \times 10^6$.

For small oscillation amplitudes the membrane is well described as a harmonic oscillator and $\omega_{cav}(x)$ is linear to lowest order in $x$ (unless the membrane is at an extremum of $\omega_{cav}(x)$). Thus this device can mimic the traditional optomechanical systems illustrated in Fig. 1(a), but without the technical challenge of integrating mirrors into cantilevers.

To illustrate this point, we use the mechanism described in Refs. [10-14] to laser cool the membrane's Brownian motion. Figure 3 shows the PSD of the membrane's undriven motion $S_x(\nu)$ when the laser is slightly red detuned from the cavity resonance. The membrane's motion is monitored via the light reflected from the cavity while the laser detuning and $P_{in}$ are varied. As described extensively elsewhere[10-14] the radiation pressure exerted by the red-detuned laser damps the membrane's Brownian motion.

We extract the membrane's effective temperature $T_{eff}$ from the data in Fig. 3 in two ways: $T_{eff}^{(x)} = m\omega_m^2 \langle x^2 \rangle / k_B$ or $T_{eff}^{(Q)} = TQ_{eff}/Q$ where $\langle x^2 \rangle = \int S_x(\nu) d\nu$ and the effective $Q$ factor $Q_{eff}$ is extracted by fitting each curve. $T_{eff}^x$ and $T_{eff}^Q$ agree to within a factor < 2. Since $T_{eff}^Q$ is insensitive to the absolute calibration of $x$, we cite $T_{eff}^Q$ in Fig. 3.

The lowest temperature achieved in Fig. 3 is 6.82 mK, a factor of $4.4 \times 10^4$ below the starting temperature of 294 K. To consider the fundamental limit to cooling, we note that the upper limit on $\beta_M$ means $F$ can be increased to > 285,000 using higher reflectivity (but commercially available) mirrors. Such a cavity should cool from $T = 300$

K to 1 mK (equivalent to $n = 200$ membrane phonons) using $P_{in} = 1$ nW or from $T = 300$ mK to the membrane's ground state using $P_{in} = 0.1$ nW[25]. Experiments with improved electronics, higher finesse mirrors, cryogenic pre-cooling, and *in situ* alignment are underway in our lab.

The laser cooling in Fig. 3 was obtained by positioning the membrane so $\omega_{cav}(x) \propto x$. However if the membrane is positioned at an extremum of $\omega_{cav}$ then $\omega_{cav}(x) \propto x^2$. In this case light leaving the cavity carries information only about $x^2$. The ability to realize a direct $x^2$-measurement is an important fundamental difference between our approach and previous work because it can be used as a quantum nondemolition (QND) readout of the membrane's phonon number eigenstate.

To see this we note that the Hamiltonian for the optomechanical system is given by $\hat{H} = \hbar\omega_{cav}(\hat{x})\hat{a}^\dagger\hat{a} + \hbar\omega_m\hat{b}^\dagger\hat{b}$ where $\hat{a}$ and $\hat{b}$ are the lowering operators for the optical and mechanical modes, $\hat{x} = x_m(\hat{b}^\dagger + \hat{b})$, and $x_m = \sqrt{\hbar/2m\omega_m}$. With the membrane at an extremum of $\omega_{cav}$ (e.g., $x = 0$), we can expand $\hat{H} \approx \hbar(\omega_{cav}(0) + \tfrac{1}{2}\omega''_{cav}(0)x_m^2(\hat{b}^\dagger + \hat{b})^2)\hat{a}^\dagger\hat{a} + \hbar\omega_m\hat{b}^\dagger\hat{b}$ where $\omega''_{cav} = \partial^2\omega_{cav}/\partial x^2$. In the rotating wave approximation (RWA) (valid in the resolved-sideband limit), this becomes $\hat{H} \approx \hbar(\omega_{cav}(0) + \omega''_{cav}(0)x_m^2(\hat{b}^\dagger\hat{b} + \tfrac{1}{2}))\hat{a}^\dagger\hat{a} + \hbar\omega_m\hat{b}^\dagger\hat{b}$. Within the RWA $[\hat{H}, \hat{b}^\dagger\hat{b}] = 0$, so the membrane's phonon number can be measured without back action. In principle $\hat{b}^\dagger\hat{b}$ can be read out by monitoring the optical cavity: it experiences a detuning-per-phonon $\Delta\omega_{cav} = x_m^2\omega''_{cav}(0)$ which can be monitored via a Pound-Drever-Hall circuit.

The presence of extrema in $\omega_{cav}(x)$ thus provides an optomechanical coupling of the form required for QND measurements of the membrane's phonon number. Whether such a measurement can be used to observe a quantum jump of the membrane depends upon whether $\Delta\omega_{cav} = (16\pi^2 cx_m^2/L\lambda^2)\sqrt{2(1-r_c)}$ can be resolved in the lifetime of a phonon number state. The shot-noise limited sensitivity of a PDH detector is[26] $S_{\omega_{cav}} = \pi^3\hbar c^3/16F^2L^2\lambda P_{in}$, and the (power) signal-to-noise ratio for resolving a jump from the $n^{th}$ phonon state is $\text{SNR}^{(n)} = \tau_{tot}^{(n)}\Delta\omega_{cav}^2/S_{\omega_{cav}}$. For realistic parameters, we find the phonon lifetime $\tau_{\text{tot}}^{(n)}$ is primarily limited by thermal excitations, with small corrections due to the RWA and the imperfect positioning of the membrane at $x = 0$. The relevant calculation is in the supplemental material.

For our estimates we assume $T = 300$ mK, that the membrane is laser-cooled to its ground state ($n = 0$), and the cooling laser is then shut off. We calculate $\text{SNR}^{(0)}$, the signal-to-noise ratio for observing the quantum jump of the membrane out of its ground state. Table 1 shows two sets of experimental parameters which give $\text{SNR}^{(0)} \sim 1$. The parameters in Table 1, though challenging, seem feasible. We have measured $Q = 1.2 \times 10^7$ for these membranes at $T = 300$ mK, and have cryogenically cooled similar devices' Brownian motion to 300 mK[27]. The membranes' low optical absorption suggests that achieving $F > 3 \times 10^5$ should be possible and that $P_{\text{in}} \sim \mu$W would not lead to excessive heating. $m = 5 \times 10^{-11}$ g is the motional mass of a 50 nm thick, 40 μm diameter membrane. Remarkably, recent work has shown that patterning such a membrane can lead to high values of $r_c$[28] and should allow for $r_c > 0.9995$.[29] The required picometer-scale placement of the membrane is within the stability and resolution of cryogenic positioning systems[30].

In conclusion, we have developed a new type of optomechanical coupling which resolves a number of the outstanding technical issues faced by previous approaches. It offers the fundamentally new feature of allowing a sensitive $x^2$ measurement which should enable measurements of the quantum jumps of mm-scale mechanical oscillators. We note this approach should make it straightforward to couple multiple mechanical devices to a single cavity mode. Stacking multiple chips like the one in Fig. 1(c) would give a self-aligned array of membranes which could be placed inside a cavity. Such a complex optomechanical system would be particularly interesting for studying entanglement between the membranes and/or using one membrane to provide a QND readout of another.

**Acknowledgments** We acknowledge funding by the NSF, the DFG NIM network, and a fellowship from the Sloane Research Foundation. We thank William Shanks for performing the microscopy and cryogenic measurements and Cheng Yang for assistance with laser cooling measurements.

**Supplementary Information** accompanies the paper on **www.nature.com/nature**.

**Author information** Reprints and permissions information is available at npg.nature.com/reprintsandpermissions. The authors declare no competing financial interests. Correspondence and requests for materials should be addressed to J.G.E.H. (jack.harris@yale.edu).

**Figure 1 | Schematic of the dispersive optomechanical set-up. a**, Conceptual illustration of "reflective" optomechanical coupling. The cavity mode (green) is defined by reflective surfaces, one of which is free to move. The cavity detuning is proportional to the displacement $x$. **b**, Conceptual illustration of the "dispersive" optomechanical coupling used in this work. The cavity is defined by rigid mirrors. The only mechanical degree of freedom is a thin dielectric membrane in the cavity mode (green). The cavity detuning is a periodic function of the displacement $x$. The total cavity length is $L = 6.7$ cm in our experiment. **c**, Photograph of a 1 mm-square, 50 nm-thick SiN membrane on a 200 μm thick Si chip. **d**, Schematic of the optical and vacuum setup. The vacuum chamber (dotted line) is ion pumped to ~ $10^{-6}$ Torr. The membrane chip is shown in orange. The optical path includes an acoustooptic modulator (AOM) for switching on and off the laser beam and a PI servo loop for locking the laser to the cavity. **e**, Calculation of the cavity frequency $\omega_{cav}(x)$ in units of $\omega_{FSR} = \pi c/L$. Each curve corresponds to a different value of the membrane reflectivity $r_c$. Extrema in $\omega_{cav}(x)$ occur when the membrane is at a node (or anti-node) of the cavity mode. Positive (negative) slope of $\omega_{cav}(x)$ indicates the light energy is stored predominantly in the right (left) half of the cavity, with radiation pressure force acting to the left (right).

**Figure 2 | Optical and mechanical characterization of the cavity. a**, Ringdown measurements of the cavity with the membrane removed (red) and in place (blue). The transmitted power $P_T$ is plotted as a function of time. The laser is switched off at 400 ns. An offset has been subtracted from the data. The exponential time constants ($\tau_M$ and $\tau_0$) fitted to the data correspond to cavity finesses $F_M = 15,200$ and $F_0 = 16,100$. **b**, Logarithmic greyscale plot of the cavity transmission as a function of laser detuning and membrane position. The two brightest curves correspond to the cavity's $TEM_{00}$ mode. They give a membrane reflectivity $r_c = 0.31$ (see Fig. 1(e)). The fainter curves are the

TEM$_{01}$ mode. **c**, Ringdown measurement of the membrane's lowest mechanical resonance. The fit gives a ringdown time $\tau = 2.67$ s, corresponding to $Q = \omega_m \tau/2 = 1.1 \times 10^6$.

**Figure 3 | Passive laser cooling of the membrane.** The power spectral density of the membrane's undriven motion is plotted for different values of the laser detuning (from top to bottom: 4.84, 2.18, 1.66, 1.00, and 0.71 cavity linewidths) and incident optical power $P_{in} = 114$ µW (except for the uppermost curve for which $P_{in} = 359$ µW. Solid lines are fits to a damped driven oscillator. The effective temperature $T_{eff}$ of the membrane is determined from $Q_{eff}$ (the effective $Q$ for each curve), and is indicated in the figure (the quoted error is the statistical error in fitting $Q_{eff}$). A broad feature is partially visible at the left of the lowest two data sets and was excluded from the fits. The noise floor in each curve results from a constant voltage noise at the detector and the detuning dependence of the volts-to-meters conversion.

**Table 1 | Parameters for observing a membrane's quantum jumps**

| $Q$ | $T$ | $F$ | $P_{in}$ | $m$ (pg) | $\omega_m/2\pi$ | $r_c$ | $x_0$ | $\lambda$ (nm) | $\tau^{(0)}$ | SNR$^{(0)}$ |
|---|---|---|---|---|---|---|---|---|---|---|
| 1.2x10$^7$ | 0.3 K | 3x10$^5$ | 10 µW | 50 | 10$^5$ Hz | 0.999 | 0.5 pm | 532 | 0.3 ms | 1.0 |
| 1.2x10$^7$ | 0.3 K | 6x10$^5$ | 1 µW | 50 | 10$^5$ Hz | 0.9999 | 0.5 pm | 532 | 0.3 ms | 4.0 |

Two sets of experimental parameters which would allow observation of an individual quantum jump from the membrane's mechanical ground state to its first excited state.

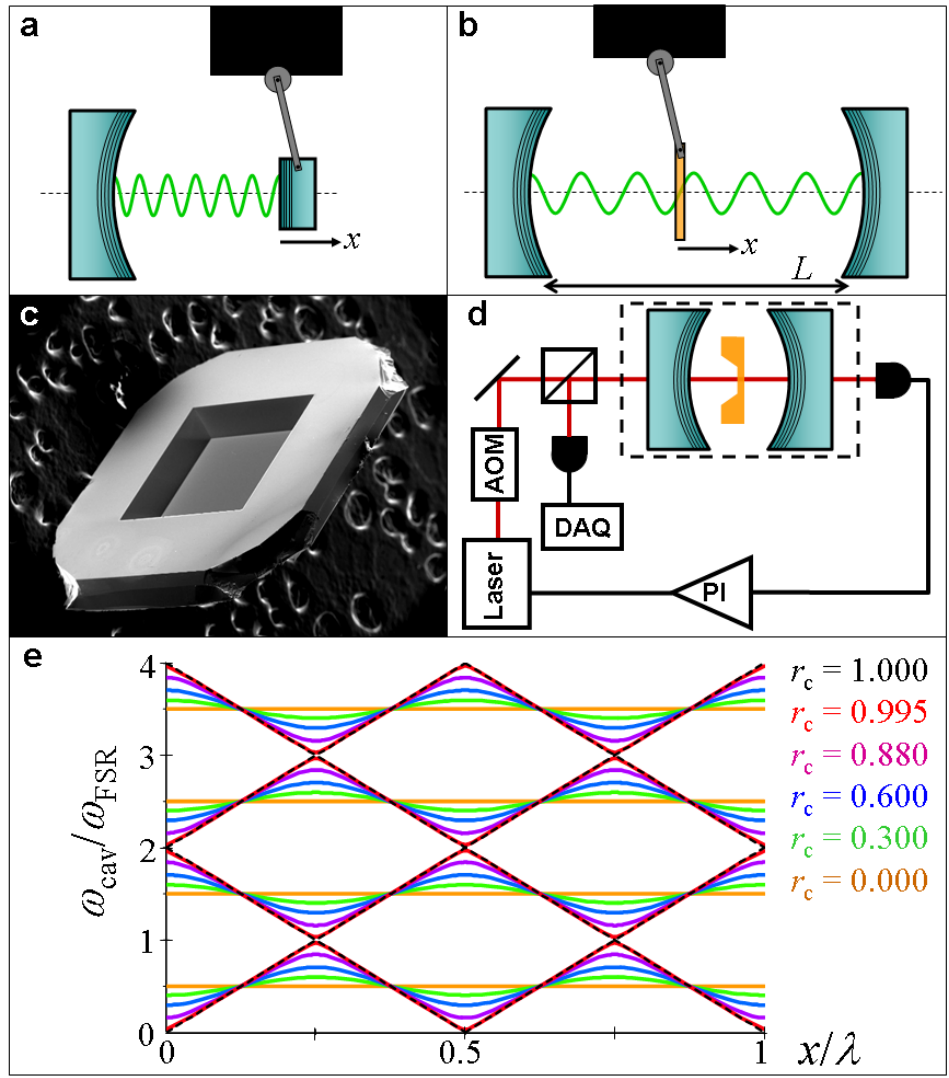

**Figure 1**

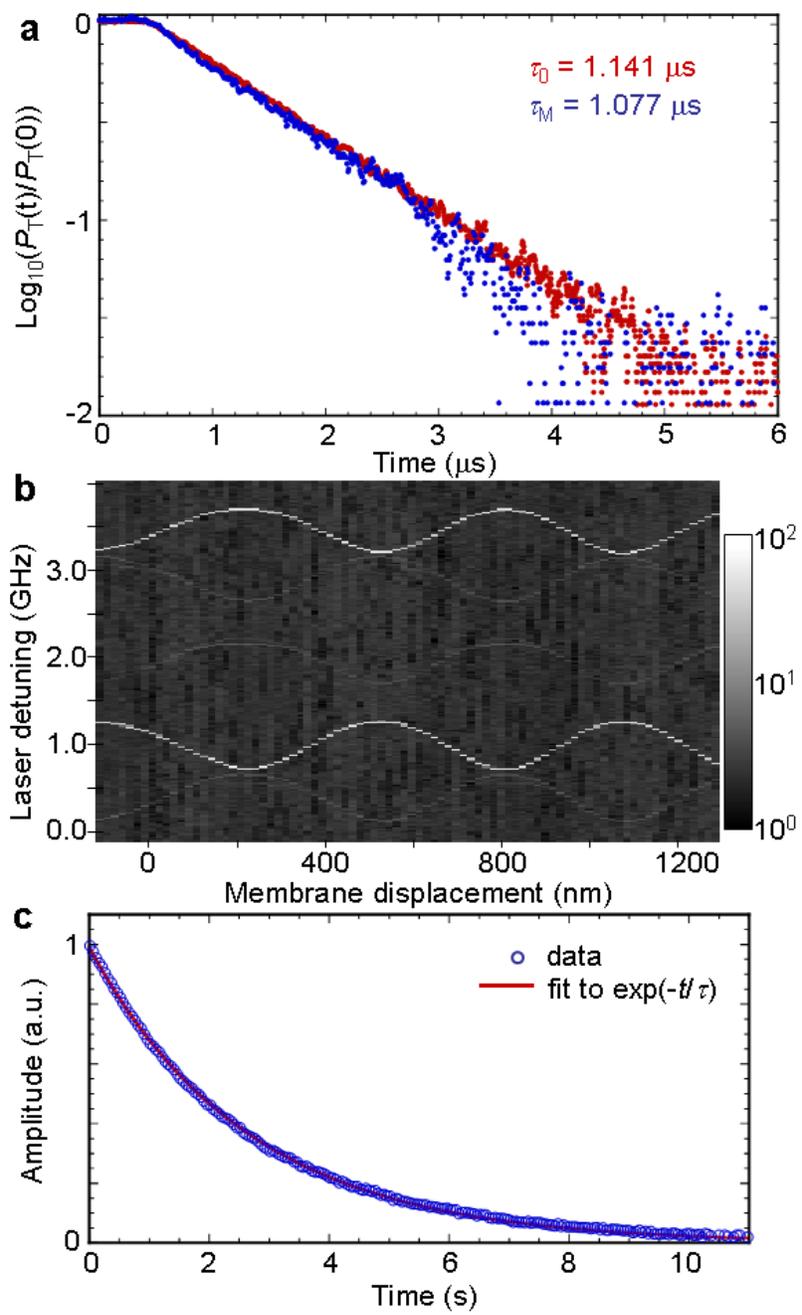

**Figure 2**

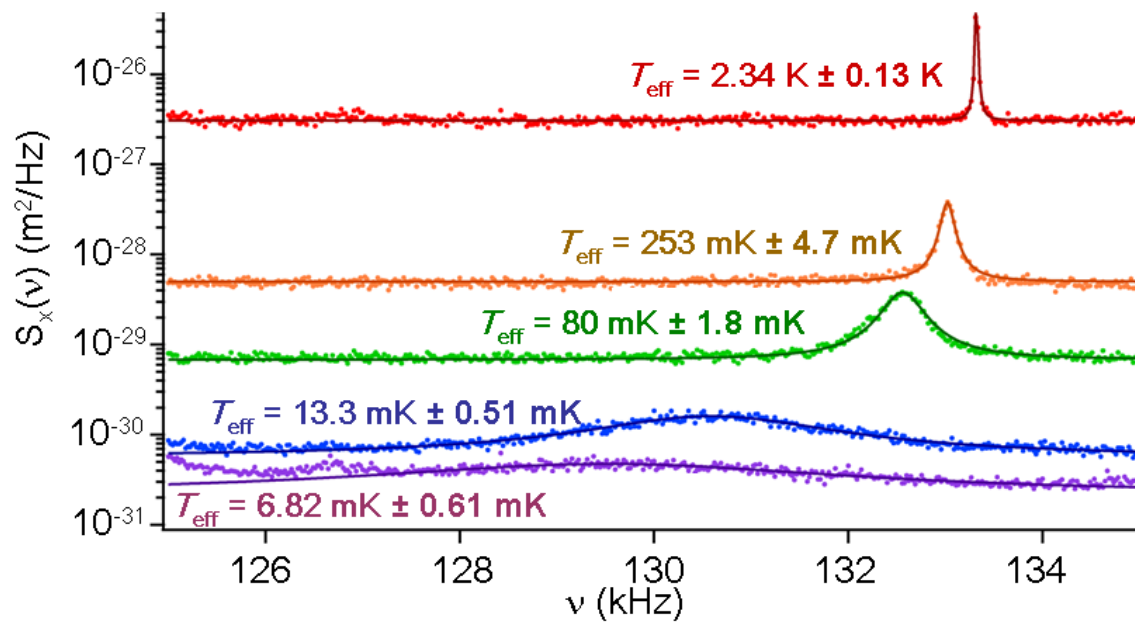

**Figure 3**

# Supplementary material for "Strong dispersive coupling of a high finesse cavity to a micromechanical membrane"

J. D. Thompson[1], B. M. Zwickl[1], A. M. Jayich[1], S. M. Girvin[1,2], Florian Marquardt[3] & J. G. E. Harris[1,2]

[1] *Department of Physics, Yale University, 217 Prospect Street, New Haven CT, 06520, USA*

[2] *Department of Applied Physics, Yale University, 15 Prospect Street, New Haven CT, 06520, USA*

[3] *Physics Department, Center for NanoScience, and Arnold Sommerfeld Center for Theoretical Physics, Ludwig Maximillians University, Theresienstrasse 37, 80333, Munich, Germany*

The purpose of this supplementary material is to estimate the feasibility of observing the quantum jumps of a micromechanical device using the setup described in our paper. To do this, we need to estimate three quantities: the cavity frequency shift per membrane phonon (i.e., the signal); the sensitivity with which the cavity frequency can be measured (i.e., the noise spectral density); and the lifetime of a phonon-number state (i.e., the allowable averaging time). Together these three quantities give the signal-to-noise ratio for observing the quantum jump. We assume throughout that the membrane has been cooled to its ground state and the cooling laser switched off.

**The shift in the cavity frequency per phonon in the membrane.** The Hamiltonian for the optomechanical device is (excluding damping and driving terms):

$$\hat{H} = \hat{N}\hbar\omega_{cav}(x) + \hat{n}\hbar\omega_m$$
$$\omega_{cav}(x) = \frac{c}{L}\cos^{-1}(r_c \cos(4\pi x/\lambda))$$
(1)

Here $\hat{N}$ is the photon number operator, $\omega_{cav}(x)$ is the cavity frequency as a function of membrane displacement, $\hat{x}$ is the membrane displacement, $\hat{n}$ is the phonon number operator, $\omega_m$ is the membrane's natural frequency and $r_c$ is its field reflectivity.

For QND measurements we want to operate near an extremum in $\omega_{cav}(x)$. Expanding about some equilibrium membrane position $x_0 \approx 0$ ($x_0$ is a constant) we have:

$$\omega_{cav}(x) \approx \omega_{cav,0} + \omega'_{cav,0}(x-x_0) + \omega''_{cav,0}(x-x_0)^2/2, \tag{2}$$

where to lowest nonvanishing order in $x_0$

$$\omega_{\gamma,0} = c\cos^{-1}(r_c)/L \tag{3}$$

$$\omega'_{\gamma,0} = \frac{16\pi^2 c r_c}{L\lambda^2 \sqrt{1-r_c^2}} x_0 \tag{4}$$

$$\omega''_{\gamma,0} = \frac{16\pi^2 c r_c}{L\lambda^2 \sqrt{1-r_c^2}} \tag{5}$$

Note that if the membrane is positioned precisely at an extremum in $\omega_{cav}(x)$ (i.e., at $x_0 = 0$) then $\omega'_{\gamma,0} = 0$ and we have just quadratic detuning.

Now we identify $(x - x_0)$ as the dynamical variable describing the membrane displacement and quantize it to become $\hat{x}$. For the present we assume $x_0 = 0$ exactly, so we can ignore the linear detuning.

Then substituting

$$\hat{x}^2 = x_m^2 (\hat{b}^\dagger + \hat{b})^2 \, , \tag{6}$$

where $x_m = \sqrt{\hbar/2m\omega_m}$ is the zero-point amplitude of the membrane, gives

$$\hat{H} = \hat{N}\hbar\left(\omega_{\gamma,0} + \tfrac{1}{2}\omega''_{\gamma,0} x_m^2 (\hat{b}^\dagger + \hat{b})^2\right) + \hat{n}\hbar\omega_m \tag{7}$$

If the cavity line is narrow enough to make the $\hat{b}^{\dagger 2}$ & $\hat{b}^2$ terms in (7) irrelevant (i.e., in the rotating wave approximation (RWA)), this becomes

$$\hat{H} = \hat{N}\hbar\left(\omega_{\gamma,0} + \Delta\omega_\gamma(\hat{b}^\dagger\hat{b} + \tfrac{1}{2})\right) + \hat{n}\hbar\omega_m \tag{8}$$

where $\Delta\omega_\gamma$ is the cavity shift per phonon. For $r_c \sim 1$ this is given by:

$$\bullet \quad \Delta\omega_\gamma = \omega''_{\gamma,0} x_m^2 = \frac{8\pi^2 c}{L\lambda^2 \sqrt{2(1-r_c)}} \frac{\hbar}{m\omega_m} \, . \tag{9}$$

**The shot-noise limited frequency resolution** of the Pound-Drever-Hall scheme leads to an angular frequency noise power spectral density[1] (i.e., in units of $s^{-2}Hz^{-1}$):

$$\bullet \quad S_\omega = \frac{\pi^3 \hbar c^3}{16 F^2 L^2 \lambda P_{in}} = \frac{\kappa}{16\bar{N}} \tag{10}$$

Where $\kappa = \pi c/LF$ is the cavity damping and $\bar{N}$ is the mean number of photons circulating in the cavity. This formula can be understood qualitatively by noting that during an observation time $t$ a number $N \propto \bar{N}\kappa t$ of photons passes through the cavity which gives a shot noise limit $\delta\theta \propto 1/\sqrt{N} \propto \delta\omega/\kappa$ for the resolvable phase shift $\delta\theta$ (or the corresponding frequency shift $\delta\omega$). The spectral density in equation (10) then follows via $S_\omega \propto \delta\omega^2 t$.

**The lifetime of a membrane phonon-number state *n*** is limited by three effects. The first is the thermal lifetime given by[2]:

$$\tau_T = \frac{Q}{\omega_m(n(\bar{n}+1)+\bar{n}(n+1))}, \quad (11)$$

where the bath's mean phonon number $\bar{n} = k_B T / \hbar \omega_m$ (we assume $k_B T / \hbar \omega_m \gg 1$). If we also assume the membrane has been laser cooled to its ground state (*n*=0) then (11) becomes:

$$\bullet \quad \tau_T = \frac{Q\hbar}{k_B T} \quad (12)$$

The second effect we consider is due to the terms discarded from $\hat{H}$ as a result of the RWA. These terms are:

$$\hat{N}\hbar\omega''_{\gamma,0}x_m^2(\hat{b}^\dagger\hat{b}^\dagger + \hat{b}\hat{b})/2 = \hat{N}\hbar\Delta\omega_\gamma(\hat{b}^\dagger\hat{b}^\dagger + \hat{b}\hat{b})/2. \quad (13)$$

Again, we assume that the membrane has been laser-cooled to its ground state. From Fermi's golden rule, the non-RWA terms will generate transitions from *n* = 0 to *n* = 2 at a rate:

$$R_{0\to 2} = \tfrac{1}{2}(\Delta\omega_\gamma)^2 S_{NN}(-2\omega_m), \quad (14)$$

where [3]

$$S_{NN}(\omega) = \int dt \exp(i\omega t) \langle \hat{N}(t)\hat{N}(0) \rangle = \bar{N} \frac{\kappa}{(\omega+\Delta)^2 + (\kappa/2)^2} \tag{15}$$

is the photon shot noise (power) spectral density in the cavity and represents the power available via Raman processes to decrease the membrane's energy (for positive $\omega$) or increase it (negative $\omega$). Here $\Delta$ is the laser detuning relative to the cavity. Since we are considering displacement detection using a Pound-Drever-Hall detector, we take $\Delta = 0$ (i.e., the probe laser locked to the cavity). Therefore

$$S_{NN}(-2\omega_m) = \bar{N} \frac{\kappa}{(2\omega_m)^2 + (\kappa/2)^2}. \tag{16}$$

This gives

$$R_{0 \to 2} = \frac{(\Delta\omega_\gamma)^2 \bar{N}}{8} \frac{\kappa}{\omega_m^2 + \kappa^2/16} \tag{17}$$

Plugging in our expression for $\Delta\omega_\gamma$ from above and using $\bar{N}\kappa = P_{in}\lambda/\pi\hbar c$ we have

$$\bullet \quad \tau_{RWA} = R_{0\to2}^{-1} = \frac{\lambda^3 L^2 (1-r_c) m\omega_m (\omega_m^2 + \kappa^2/16)}{8\pi^3 x_m^2 c P_{in}} \tag{18}$$

Lastly, there is the excitation rate due to the membrane not being exactly at the extremum of the band structure. This adds the following term to the Hamiltonian

$$\hat{N}\hbar\omega'_{\gamma,0} x_m (\hat{b}^\dagger + \hat{b}). \tag{19}$$

Again, using Fermi's golden rule, this will generate transitions out of the membrane's ground state at a rate

$$R_{0 \to 1} = (\omega'_{\gamma,0} x_m)^2 S_{NN}(-\omega_m) \tag{20}$$

Using (15), we get:

- $$\tau_{lin} = R_{0 \to 1}^{-1} = \frac{m \omega_m L^2 \lambda^3 (1-r_c)(4\omega_m^2 + \kappa^2)}{256 \pi^3 P_{in} c x_0^2} \tag{21}$$

The total lifetime of the membrane's ground state is then

- $$\tau^{(0)} = 1/(\tau_T^{-1} + \tau_{RWA}^{-1} + \tau_{lin}^{-1}) \tag{22}$$

**The signal-to-noise ratio** for observing a single quantum jump out of the membrane's ground state is then**:**

- $$SNR^{(0)} = (\Delta \omega_\gamma)^2 \tau^{(0)} / S_\omega \tag{23}$$

Finally, we note that the different contributions to the lifetime obey the following relations (in the good cavity regime, where $\omega_m \gg \kappa$, which is the most relevant regime). The ratio of the total lifetime to the lifetime generated by a finite displacement is given by

$$\frac{\tau^{(0)}}{\tau_{lin}} = \frac{SNR^{(0)}}{16} \left(\frac{x_0}{x_m}\right)^2 \left(\frac{\kappa}{\omega_m}\right)^2 \tag{24}$$

For the parameters used in the numerical estimates in our main paper (e.g., Table 1), the total lifetime is dominated by thermal transition and the ratio in (24) is small. Furthermore, the lifetime correction related to non-RWA effects is even smaller, since

$$\frac{\tau_{\text{lin}}}{\tau_{\text{RWA}}} = \frac{1}{8}\left(\frac{x_m}{x_o}\right)^2 \qquad (25)$$

is much smaller than unity for reasonable estimates of the positioning accuracy $x_0$.

The parameters in Table 1 should readily allow for laser cooling to the membrane's ground state, as is assumed in these calculations. In addition, these parameters satisfy the condition $\tau^{(0)} > 1/\omega_m$, necessary for a QND measurement. Lastly, we note that when $r_c$ approaches unity and $x_0$ approaches 0, two cavity modes approach degeneracy (as can be seen in Fig. 1(e) of our main paper). For our analysis to be valid, the gap $\Delta_{\text{gap}}$ between these two modes should be greater than $\omega_m$. Since $\Delta_{\text{gap}} \approx (c/L)\sqrt{8(1-r_c)}$, this conditioned is satisfied as long as $1-r_c > 10^{-8}$.

Even if individual quantum jumps cannot easily be resolved, the approach outlined here can still be used to observe energy quantization in the membrane. Repeated measurements of the type described here (even with SNR < 1) could be converted to histograms and averaged together to reveal discretization of the membrane's energy. While less dramatic than observations of individual quantum jumps, such an observation would still represent a major breakthrough.